**IceCube: Neutrino Physics from GeV-PeV**

Francis Halzen for the IceCube Collaboration

IceCube transforms approximately one cubic kilometer of natural Antarctic ice into a Cherenkov detector; see Fig.1. The IceCube detector collects approximately 100,000 atmospheric neutrinos per year in the 0.1–100 TeV energy range. Both the high statistics and the higher energy range represent opportunities for particle physics discovery. Measurements of the atmospheric muon and electron neutrino spectrum[1] have been extended to 100 TeV; atmospheric events above this energy are extremely rare.

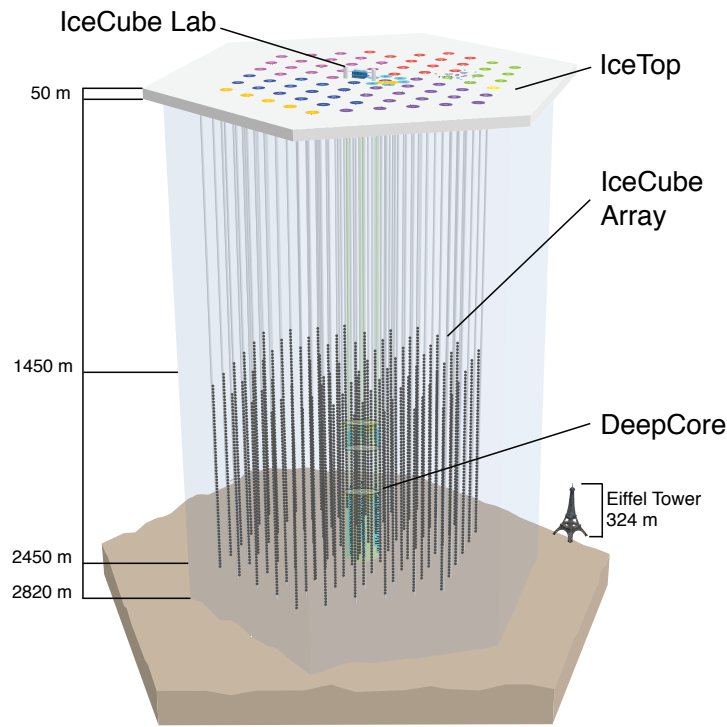

*Figure 1. IceCube instruments a cubic kilometer of Antarctic glacial ice. It detects neutrinos by observing Cherenkov light from secondary charged particles produced in neutrino-nucleon interactions. This light is detected by an array of 5160 digital optical modules (DOMs), each of which contains a photomultiplier and readout electronics housed in a glass pressure sphere. The DOMs are arranged into an array of 86 vertical strings, with 60 DOMs per string at depths between 1450 m and 2450 m. Outside of the DeepCore low-energy subarray, these DOMs are vertically spaced at 17-meter intervals and the strings are on average 125 m apart horizontally. The DeepCore subarray fills in the center of the detector with a denser array of photomultipliers and provides a lower energy threshold of 10 GeV over a substantial fraction of the IceCube volume.*

At the same time, two events with energy exceeding 1 PeV were discovered serendipitously in a search for cosmogenic neutrinos[2]. They are showers initiated by electron or tau neutrinos contained inside the detector volume. To trace their origin, an all-sky search was performed, using data collected between May 2010 and May 2012, for neutrino events with energies above approximately 50 TeV and with vertices contained in the detector. Twenty-eight events of this type were observed. These events represent the highest energy sample of neutrinos ever observed[3]. Their flavors, arrival directions, and energies are consistent with generic expectations for neutrinos of extraterrestrial origin; see Fig.2.

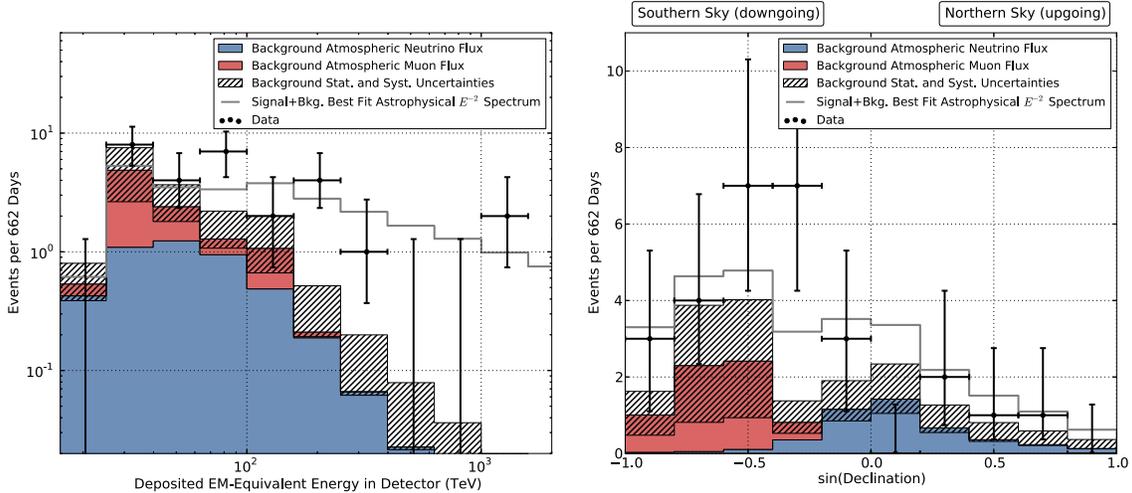

*Figure 2. Distribution of the deposited energies (left) and declination angles (right) of the observed events compared to model predictions. Energies plotted are in-detector visible energies, which are lower limits on the neutrino energy. Note that deposited energy spectra are always harder than the spectrum of the neutrinos that produced them due to the neutrino cross-section increasing with energy. The expected rate of atmospheric neutrinos is based on northern hemisphere muon neutrino observations. The estimated distribution of the background from atmospheric muons is shown in red. Due to lack of statistics from data far above our cut threshold, the shape of the distributions from muons in this figure has been determined using Monte Carlo simulations with total rate normalized to the estimate obtained from our in-data control sample. Combined statistical and systematic uncertainties on the sum of backgrounds are indicated with a hatched area. The gray line shows the best-fit $E^{-2}$ astrophysical spectrum with all-flavor normalization (1:1:1) of $E^2\Phi_v(E) = 3.6 \cdot 10^{-8}$ GeV cm$^{-2}$ s$^{-1}$ sr$^{-1}$ and a spectral cutoff of 2 PeV.*

An extraterrestrial beam of neutrinos provides new opportunities for neutrino physics, among them the study of neutrino oscillations over cosmic baselines and the possibility to probe the quantum structure of spacetime. The small perturbation by Planckian physics on the propagation of a neutrino becomes observable when integrated over cosmic distances. With energies of 100 TeV and masses of order $10^{-2}$ eV or less, even the atmospheric neutrinos observed by IceCube reach Lorentz factors of $10^{16}$, possibly larger.

With the low-energy extension DeepCore, IceCube's threshold has been lowered to 10 GeV over a significant effective volume. Using conventional IceCube analysis tools, neutrino oscillations have been observed above the 5σ level[4]; the development of methods specialized for low energy should result in a competitive precision measurement of the oscillation parameters. More importantly, IceCube studies oscillations at energies that exceed those of present experiments by one order of magnitude. This represents an opportunity for a highly sensitive search for any new physics that interfere with the standard oscillation pattern[5].

At TeV energy, the sensitivity of IceCube data to sterile neutrinos in the eV mass range potentially exceeds that of any other experiment[6] and is only limited by systematic errors[7]. A dedicated analysis is underway.

IceCube will observe the next Galactic supernova, collecting over one million neutrinos in millisecond time bins from an explosion at the center of the Galaxy. The data has the potential to reveal a wealth of neutrino physics, including the neutrino mass hierarchy[8].

Finally, IceCube is a novel instrument with the potential of discovery. Looking for the "unexpected" includes searching for particle emission from cosmic strings or any other form of topological defects or heavy cosmological remnants created in the early Universe, for magnetic monopoles[9], Q-balls and the like.

**References:**

1. M.G. Aartsen et al. [IceCube Collaboration], Phys. Rev. Lett. 110 (2013) 151105 [arXiv:1212.4760 [hep-ex]]; R. Abbasi et al. [IceCube Collaboration], Phys. Rev. D 84, 082001 (2011) [arXiv:1104.5187 [astro-ph.HE]]; N. Whitehorn, Snowmass 2013, https://indico.fnal.gov/getFile.py/access?contribId=182&sessionId=43&resId=0&materialId=slides&confId=6890.

2. M.G. Aartsen et al. [IceCube Collaboration], Phys. Rev. Lett. 111, 021103 (2013) [arXiv:1304.5356 [astro-ph.HE]].

3. M.G. Aartsen et al. [IceCube Collaboration], submitted to Science.

4. M.G. Aartsen et al. [IceCube Collaboration], submitted to Phys. Rev. Lett. [arXiv:1305.3909 [astro-ph.HE]].

5. R. Abbasi et al. [IceCube Collaboration], Phys. Rev. D 82, 112003 (2010) [arXiv:1010.4096 [astro-ph.HE]].

6. A. Esmaili, F. Halzen and O.L.G. Peres, JCAP 1211, 041 (2012) [arXiv:1206.6903 [hep-ph]].

7. F. Halzen for the IceCube Collaboration, arXiv:1111.0918 [hep-ph].

8. R. Abbasi et al. [IceCube Collaboration], Astron. Astrophys. 535, A109 (2011) [arXiv:1108.0171 [astro-ph.HE]].

9. R. Abbasi et al. [IceCube Collaboration], Phys. Rev. D87 (2013) 022001 [arXiv:1208.4861 [astro-ph.HE]]